\pdfoutput = 1
\documentclass[twocolumn,showpacs,prl,amsmath,amssymb,floatfix]{revtex4-1}

\usepackage[dvipsnames]{xcolor}

\usepackage{bm}
\usepackage{epsfig}
\usepackage{psfrag}
\usepackage[english]{babel}
\usepackage{amsfonts}
\DeclareSymbolFont{bbold}{U}{bbold}{m}{n}
\DeclareSymbolFontAlphabet{\mathbbold}{bbold}

\renewcommand{\Im}{\mathop{\text{Im}}}
\renewcommand{\Re}{\mathop{\text{Re}}}
\newcommand{\up}{\uparrow}
\newcommand{\down}{\downarrow}
\newcommand{\eps}{\varepsilon}

\begin{document}

\title{Evolution of the transmission phase through a Coulomb-blockaded Majorana wire}

\author{Casper Drukier, Heinrich-Gregor Zirnstein, and Bernd Rosenow}
\affiliation{Institut f\"{u}r Theoretische Physik, Universit\"{a}t
  Leipzig,  Br\"{u}derstrasse 16, 04103 Leipzig, Germany}

\author{Ady Stern and Yuval Oreg}
\affiliation{Department of Condensed Matter Physics, Weizmann Institute of Science, Rehovot, 76100, Israel}

\date{\today}

\begin{abstract}
We present a study of the transmission of electrons through a semiconductor quantum wire with strong spin-orbit coupling in proximity to an s-wave superconductor, which is Coulomb-blockaded. Such a system supports Majorana zero modes in the presence of an external magnetic field. Without superconductivity, phase lapses are expected to occur in the transmission phase, and we find that they disappear when a topological superconducting phase is stabilized. We express tunneling through the nanowire with the help of effective matrix elements, which depend on both the fermion parity of the wire and the overlap with Bogoliubov-de-Gennes wave functions. Using a modified scattering matrix formalism, that allows for including electron-electron interactions, we study the transmission phase in different regimes.
\end{abstract}


\maketitle

\emph{Introduction.}
Majorana zero modes (MZMs) are localized zero energy states that can arise in topological superconductors \cite{Alicea12,Beenakker13}. In the last decade, they have attracted much attention, because they are promising candidates for the realization of quantum computation~\cite{Kitaev01,Hyart13,Aasen15}. Recent progress \cite{Mourik12,Rohkinson12,Deng12,Churchill13,Das12,Finck13,Albrecht16,Fadaly17,Suominen17} suggests that MZMs can be realized experimentally, and that they can be detected by electric conductance measurements.

The motion of electrons through a mesoscopic device is characterized by a transmission matrix $T$. In the simplest case, it reduces to a complex number. From the Landauer formula the conductance is proportional to the square of its absolute value. Its quantum mechanical phase determines how electrons moving along different trajectories interfere. A typical interference experiment consists of two arms that electrons can travel through. One containing the device, the other one acting as a reference arm. When the arms join, the electrons interfere due to different relative phases. The phase in the reference arm can be adjusted by means of the Aharonov-Bohm effect, leading to oscillations of the total conductance depending on a magnetic flux within the interferometer loop.\cite{Oreg92,Yacoby95,Schuster97,Hackenbroich97,Taniguchi99,Baltin99,Ji00,Yeyati00,Silvestrov00,Silva02,EntinWohlman02,Aharony02,Aharony03,Kalish05,Apel05,Berkovits05,Meden06,Golosov06,Oreg07,Karrasch07,Goldstein07,Dinaii14}

\begin{figure}[h]
\begin{center}
    \includegraphics[width=\linewidth]{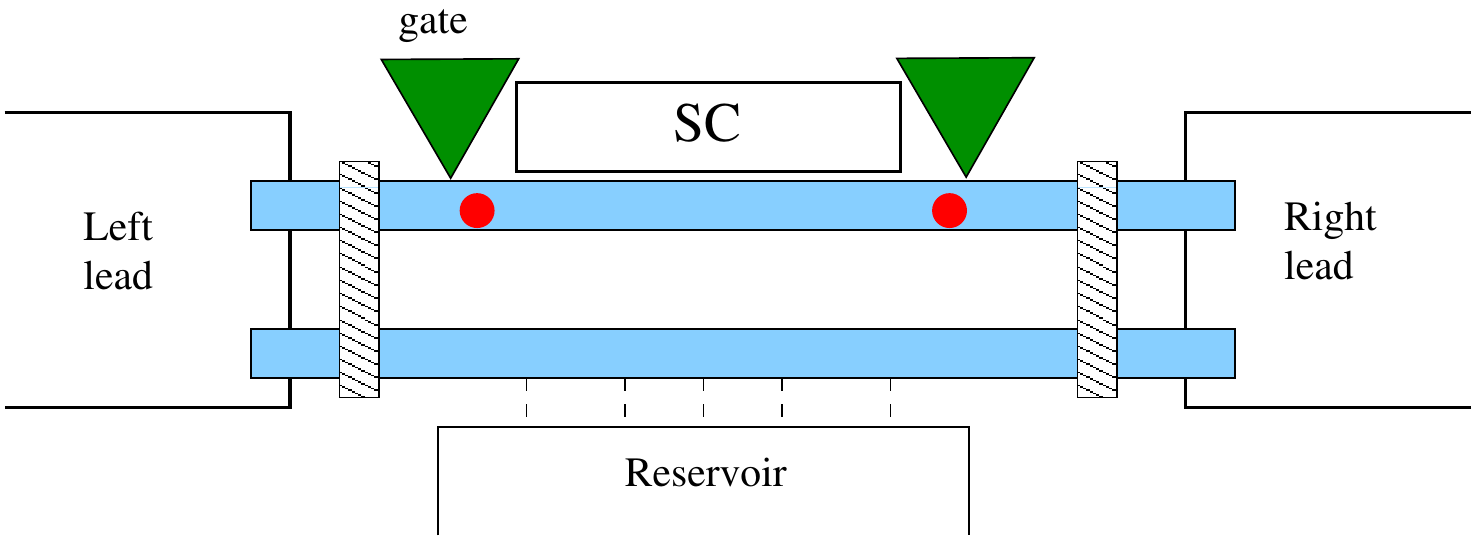}
\end{center}
\caption{Schematic setup. We consider two wires (azure rectangles) connected to leads to the left and right. The shaded rectangles denote tunneling bridges, and the triangles gates which control the coupling of the Majorana to the leads. The upper wire is in proximity to a conventional s-wave superconductor (SC in the figure). The red circles represent Majorana zero modes. We couple the reference arm of the interferometer to a reservoir to avoid the phase rigidity effect (coupling indicated by dashed lines)\cite{Aharony02}.}
\label{fig:setup}
\end{figure}

In this paper, we study the transmission phase when the device is a quantum dot made of a topological superconductor that can host MZMs~\cite{Read00,Kitaev01,Fu08,Sau10,Lutchyn10,Oreg10,Alicea11}.
In particular, in the Coulomb blockade regime, it has been predicted that the transmission phase is sensitive to the presence of MZMs~\cite{Fu10,Landau16,Plugge16a,Vijay16,Plugge16b,Hansen18}.
For concreteness, we consider a semiconducting nanowire with Rashba-type spin-orbit coupling covered by a metallic superconductor such as Aluminum~\cite{Lutchyn10,Oreg10,Alicea11}, see Fig.~\ref{fig:setup}.

We find that the trivial and topological regime can be distinguished by the presence or absence of phase lapses, where the phase exhibits an abrupt change of $\pi$. The presence/absence of phase lapses may be understood as originating from the spatial symmetry of Bogoliubov-de-Gennes (BdG) wave function amplitudes and the way they evolve as we scan through consecutive Coulomb blockade peaks. In neighboring peaks, the dominant contribution to the transmission amplitude switches between being electron type and hole type.  In the topological regime,  the spatial symmetry of the effective p-wave pairing leads to an  opposing inversion symmetry of these amplitudes and thus absence of phase lapses. In contrast, in the non-topological regime, the doubling of the number of Fermi points invalidates this argument, and hence may introduce phase lapses. We point out that this depends on the spin polarization and discuss how the transmission phase is influenced by external parameters, with and without breaking an effective time reversal symmetry that may occur in these wires. We also explicitly distinguish between cases where consecutive Coulomb blockade peaks are dominated by tunneling through the same level or through consecutive levels.

In the Coulomb blockade regime, the transmission depends on the total number $N_0$ of electrons in the quantum dot, which is comprised of the semiconducting wire and the superconducting coating.
If the dot is capacitively coupled to a gate with voltage $V_G$, then its energy includes a charging term $H_c = E_c N_0^2/2 - eV_G N_0$. At small bias voltage, an electron may only tunnel if there is no energy cost for allowing the electron into the dot, i.e.~if the charging terms for $N_0$ and $N_0+1$ are roughly equal. This occurs when the gate voltage $eV_G$ takes certain discrete values $E_{N_0}$.
As a function of gate voltage, the transmission $T(V_G)$ has sharp conductance peaks at these values. By considering only two resonances, and by assuming them to be  independent of each other, these resonances can be described by a Breit-Wigner form obtained from the retarded Green function of the wire \cite{Meir92}. Between two resonances, $E_{N_0} < eV_G < E_{N_0+1}$, we find, taking into
account the lack of degeneracy of the quantum dot spectrum (for details and limits of applicability see \cite{supplement}).

\begin{widetext}
\begin{align}
\label{eq:breit-wigner}
T_{\sigma\sigma}(V_G) = \sum_{\substack{N=N_0, \\N_0+1}}
\frac{\rho_F\lambda_{\sigma L}(N)\lambda_{\sigma R}^\ast(N)}{eV_G-E_{N} + i\pi\rho_F\sum_{\sigma^\prime}(|\lambda_{\sigma^\prime L}(N)|^2+|\lambda_{\sigma^\prime R}(N)|^2)}+\mathcal{O}\left(\frac{\rho_F^2\lambda^4}{\left(eV_G\right)^2}\right),
\end{align}
\end{widetext}
with $\lambda\sim\lambda_{\sigma,L/R}$. Here, $\sigma$ denotes the spin of the transmitted electron, and we 
do not consider spin-flip processes as they do not contribute to interference.
We will assume tunneling proceeds through Bogoliubov quasiparticles. The resonances then occur at the effective single-particle energy $E_{N_0}=N_0E_c+\epsilon_{n_{\textrm{min}}}$, with $\epsilon_{n_{\textrm{min}}}$ denoting the lowest Bogoliubov quasiparticle energy. This captures the behavior of the system near a resonance.
If we assume that the incoming electron tunnels via a single state in the wire, then the complex quantities $\lambda_{\sigma L}(N)$ and $\lambda_{\sigma R}(N)$ can be identified with the coupling of this state to the left and the right lead. Finally, $\rho_F$ is the density of states at the Fermi energy in the leads. For a given gate voltage, we always include the two neighboring resonances with
$E_{N_0} < e V_G < E_{N_0+1}$.

When the gate voltage $V_G$ is swept across a resonance, the phase of the transmission changes by $\pi$ according to Eq.~(\ref{eq:breit-wigner}). However, when increasing the voltage further, towards the next resonance, a phase lapse has often been seen in many interferometer studies over the years \cite{Hackenbroich97,Baltin99,Baltin99b,Silvestrov00,Berkovits05,Koenig05,Goldstein07,Molina12}. In particular, this will happen when two subsequent resonances have coefficients with equal phase, $\arg(\lambda_{\sigma L}(N)\lambda^*_{\sigma R}(N)) = \arg(\lambda_{\sigma L}(N+1)\lambda^*_{\sigma R}(N+1))$, as
the denominators in Eq.~(\ref{eq:breit-wigner}) differ by a relative minus sign. On the other hand, phase lapses are absent when the sign of the coefficients changes from one resonance to the other, such that $\arg(\lambda_{\sigma L}(N)\lambda^*_{\sigma R}(N)) - \arg(\lambda_{\sigma L}(N+1)\lambda^*_{\sigma R}(N+1)) = \pm\pi$ \cite{lapse_detection}.

In the following, we study the evolution of the transmission phase for different regimes of the nanowire system. The interference setup is illustrated in Fig.~\ref{fig:setup}. To avoid the phase rigidity effect in the presence of time-reversal symmetry \cite{Aharony02}, we include a reservoir in the setup.

\emph{Model.} The electrons in the wire are described by the BdG Hamiltonian
\begin{align}
\nonumber
\mathcal{H} &=
\left[-\frac{\hbar^2}{2m}\partial_y^2-\mu-i\alpha_R\sigma_x\partial_y\right]\tau_z
\\ &\qquad
- B_z\sigma_z - B_x\sigma_x + \Delta\tau_x,
\label{eq:wireBdG}
\end{align}
with second quantized form $\hat H_{\textrm{D}} = \frac12 \int dy\, \Psi^\dagger \mathcal{H} \Psi$, where $\Psi=(\psi_\up,\psi_\down,\psi_\down^\dagger,-\psi_\up^\dagger)$ is a Nambu vector of electron operators. The coefficient $\alpha_R$ denotes a Rashba-type spin-orbit coupling while $B_x$  and $B_z$ are Zeeman energies arising from a magnetic field in the $xz$-plane.
The proximity to the s-wave superconductor induces a pairing energy $\Delta$, which we choose to be real. Finally, the chemical potential is denoted by $\mu$, and the Pauli matrices $\sigma_i$ and $\tau_i$ act in spin- and particle-hole space respectively. Particle-hole symmetry $\{\mathcal{H},P\}=0$ is described by $P=\sigma_y\tau_y K$ where $K$ denotes complex conjugation. Additionally, we have an anti-unitary reflection symmetry $[\mathcal{H},\tilde\Pi]=0$ defined on BdG wave functions, $\Phi$, as $(\tilde\Pi\Phi)(y)=K\Phi(L-y)$ where $L$ is the length of the wire. For $B_x=0$, the Hamiltonian also has an effective time-reversal symmetry $\tilde{T}=\sigma_z K$ with $\tilde{T}^2=+1$ and $[\mathcal{H},\tilde{T}]=0$, which puts it into the BDI symmetry class of topological superconductors \cite{Altland97}. For suitable parameters, the wire hosts MZMs~\cite{Lutchyn10,Oreg10,Alicea11}.

\begin{figure*}
\includegraphics[width=\linewidth]{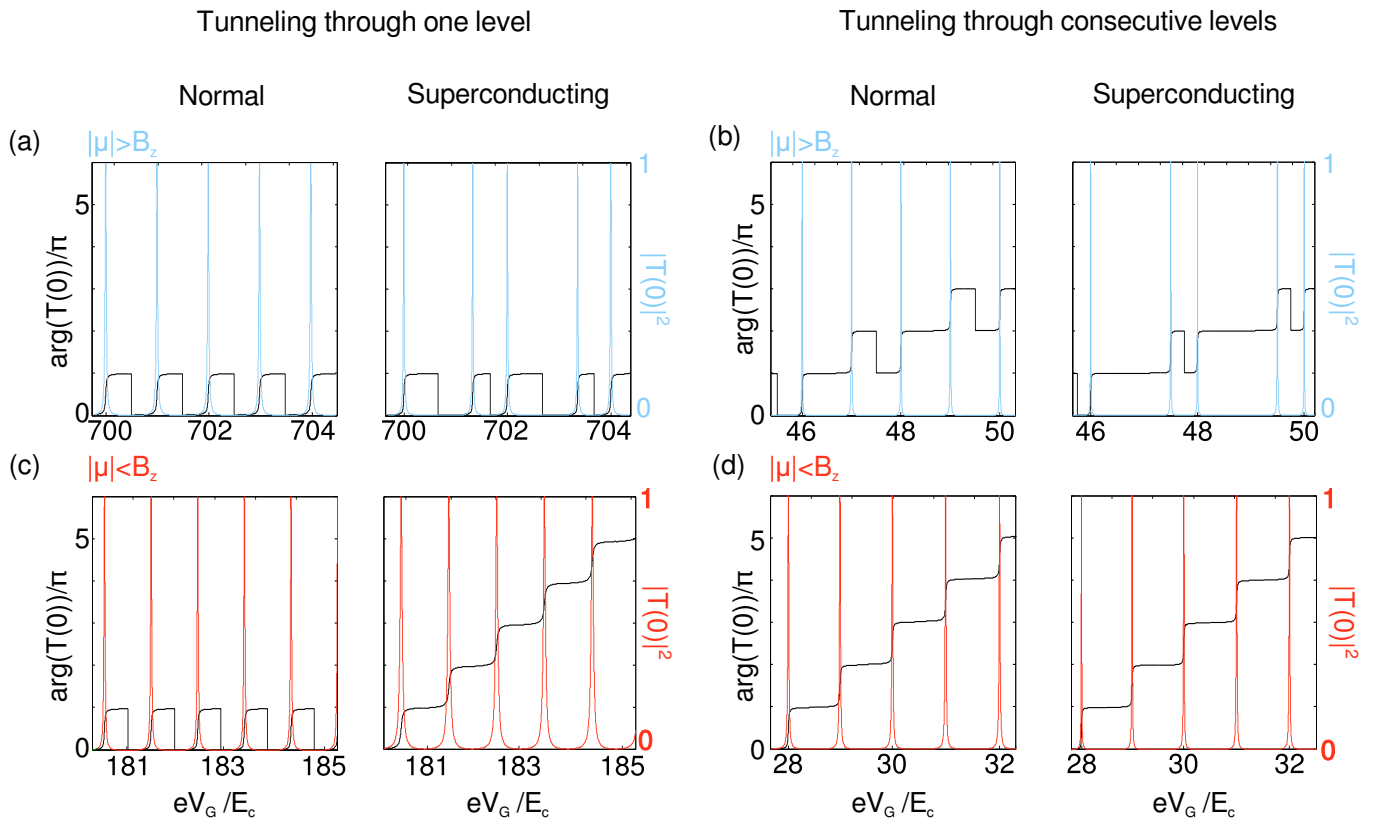}
\caption{
Plots of the the transmission phase $\textrm{arg}(T_{\up\up} + T_{\down\down})/\pi$ (black) and amplitudes $|T_{\up\up}|^2 + |T_{\down\down}|^2$ (blue and red), at zero temperature and at the Fermi level calculated using the S-matrix, for a quantum wire with tunneling through the same level in consecutive peaks [panels (a) and (c)], and tunneling through consecutive levels in consecutive peaks [panels (b) and (d)]. For the former $E_{\textrm{M}}/E_{\textrm{w}}=0.05$, such that only every 20th electron entering the hybrid system of wire and superconductor, enters the wire, whilst for the latter $E_{\textrm{M}}/E_{\textrm{w}}=1$. In the presence of a superconducting gap $\Delta$, the wire can either enter a trivial [(a) and (b)] or a topological [(c) and (d)] phase, depending on the number of particles $N_{\textrm{w}}$ in the wire.
All plots are calculated by discretizing the BdG Hamiltonian (\ref{eq:wireBdG}) on a 1D lattice with $500$ sites with parameters $L=2.5\,\mu\textrm{m}$, $m=0.02m_e$, $\tilde{t}=\hbar^2/(2ma^2)=80\,\textrm{meV}$, $u_0=\alpha_R/a=0.05\tilde{t}$, $B_z=0.003\tilde{t}$, $B_x=0$ and $E_c=3\Delta$. Here, $a$ is the lattice spacing, $\tilde{t}$ the hopping and $u_0$ an effective spin-orbit coupling in the discretized model. The two columns in each panel are results for the normal regime ($\Delta=0$) and the superconducting regime where $\Delta=2.5\delta_F$. The tunneling matrix elements $t$ on the left and right sides have been chosen to be equal and satisfy $\rho_F t^2=0.2\Delta$. The plots have been made by summing the transmission over the spin configurations. The transmission phases have been shifted by integer multiples of $\pi$ for clarity.}
\label{fig:results}
\end{figure*}

The spatial form  of the MZM wave function depends on the number of particles in the nanowire $ N_{\textrm{w}}$ via the chemical potential $\mu$. However, the total number of particles $ N_0$ in the dot is larger than $ N_{\textrm{w}}$, because electrons may also reside in the superconductor. In general, the semiconductor wire has a much larger level spacing than the superconductor, such that, at equilibrium most of the electron density will be accommodated in the superconductor. Hence the tunneling from the leads to the hybrid superconductor-wire system is through the same wire level for several subsequent resonances. Thus, tunneling amplitudes are mainly determined by wave functions in the wire, but the charge of additional Cooper pairs  mostly counts towards the charge in the superconductor, not the wire. We model this with a charging term
\begin{align}
\label{eq:chargingham}
\hat H_{c,0}=\frac{1}{2}E_0 {\hat N}_0^2+\frac{1}{2}E_{\textrm{w}}\hat N_{\textrm{w}}^2-eV_G \hat N_0-E_{\textrm{M}} {\hat N}_0 \hat N_{\textrm{w}}.
\end{align}
Here, the charging energy $E_0$ refers to the whole dot, $E_{\textrm{w}}$ to the wire, and $E_{\textrm{M}}$ denotes the coupling between them. Replacing $\hat N_{\textrm{w}}$ by $N_{\textrm{w}}=\langle\hat N_{\textrm{w}}\rangle$, and minimizing the Hamiltonian with respect to this expectation value, we obtain $N_{\textrm{w}}=(E_{\textrm{M}}/E_{\textrm{w}})\langle N_0\rangle$ and the effective charging Hamiltonian $\hat H_c$ mentioned before Eq. (\ref{eq:breit-wigner}) with $E_c=E_0-E_{\textrm{M}}^2/E_{\textrm{w}}$. In this way, we can also apply our model when tunneling occurs through subsequent wire levels.
We refer to the regime $E_{\textrm{M}}/E_{\textrm{w}} \ll 1$ as tunneling through the same level in consecutive peaks, and to the regime $\hat N_{\textrm{w}}=\hat N_0$ as tunneling through consecutive levels in consecutive peaks.

In order to calculate the transmission phase numerically, we discretize the Hamiltonian $\hat H_{\textrm{D}}$ (defined below Eq.~(\ref{eq:wireBdG})) on a lattice with hard-wall boundary conditions. The BdG equations that emerge from $H_{\textrm{D}}$ can then be solved for a given chemical potential which is determined self-consistently for a given total number of particles, and number of particles in the wire \cite{supplement}. The BdG wave functions thus obtained are the ones we use to deduce the tunneling amplitudes. For this, we supplement the dot Hamiltonian $\hat{H}_D$  with a description of the leads in terms of $\hat H_{\textrm{L}} = \sum_{\substack{k,\sigma \\ \alpha=L,R}}\epsilon_{k\alpha}c_{k\sigma\alpha}^\dagger c_{k\sigma\alpha}$, where $\alpha$ labels the left and right leads respectively, $\sigma$ denotes spin, and $c_{k\sigma\alpha}$ ($c_{k\sigma\alpha}^\dagger$) annihilates (creates) an electron in lead $\alpha$. The tunnel coupling between leads and dot is described by
\begin{align}
\hat H_{\textrm{T}} = \sum_{\substack{m\sigma \\ \alpha=L,R}}t_{\alpha m\sigma}c_\sigma^\dagger(y_\alpha)d_{m}+{\textrm{h.c.}}, \label{eq:tunhamorig}
\end{align}
where  $c_\sigma(y_\alpha)$, $c_\sigma^\dagger(y_\alpha)$ are lead operators evaluated in the vicinity of lead $\alpha$, $d_{m}$ ($d_{m}^\dagger$) annihilates (creates) an electron in the $m$'th wire state with energy $\epsilon_m$, and $t_{\alpha m\sigma}$ is the tunneling matrix element between lead $\alpha$ and the $m$'th wire state. The tunneling matrix elements are identical in magnitude, while their phase is set to be the phase of $\varphi_{m\sigma}(y)$, the eigenstates of the wire part $\hat{H}_D$ for $\Delta=0$, close to the ends of the wire.

The wire operators $d_{m}$ can now be expressed in terms of BdG operators $\beta_n$ and $\beta_n^\dagger$. Assuming the gap in the superconductor is greater than in the wire, an unpaired electron must enter the wire, occupying the lowest BdG quasiparticle state for $N_0$ odd. Denoting this state by $n_{\textrm{min}}$ we thus write $2\beta_{n_{\textrm{min}}}^\dagger\beta_{n_{\textrm{min}}}-1=(-1)^{N_0}$ in the ground state. Higher quasiparticle states are not occupied in the ground state so $\beta_n^\dagger\beta_n=0$ for $n\neq n_{\textrm{min}}$. Following Ref.~\cite{Fu10}, we project the Hamiltonian to the states with $N_0$ and $N_0+1$ particles and introduce new fermionic operators $f_n$. The tunneling part then becomes \cite{supplement}
\begin{align}
H_{\textrm{T}}=\frac{1}{2}\sum_{\substack{\alpha=L,R \\ n\sigma}}\lambda_{n\sigma\alpha}c_\sigma^\dagger(y_\alpha)f_n+\textrm{h.c.},
\label{eq:efftun}
\end{align}
with effective tunneling matrix elements
\begin{align}
\nonumber
\lambda_{n_{\textrm{min}}\sigma\alpha}
&=
\lambda_{n_{\textrm{min}}\sigma\alpha}^u+\lambda_{n_{\textrm{min}}\sigma\alpha}^v
\\ &\qquad - (-1)^{N_0}\left[\lambda_{n_{\textrm{min}}\sigma\alpha}^u-\lambda_{n_{\textrm{min}}\sigma\alpha}^v\right]
,\end{align}
and $\lambda_{n\sigma\alpha}=2\lambda_{n\sigma\alpha}^u$ for $n\neq n_{\textrm{min}}$. Here
\begin{align}
\label{eq:overlapu} \lambda_{n\sigma\alpha}^u &= \sum_m t_{\alpha m\sigma}\int dy\, \varphi_{m\sigma}^\ast(y)u_{n\sigma}(y), \\
\label{eq:overlapv} \lambda_{n\sigma\alpha}^v &= \sum_m t_{\alpha m\sigma}\int dy\, \varphi_{m\sigma}^\ast(y)v_{n\sigma}^\ast(y),
\end{align}
where $u_{n\sigma}(y)$ and $v_{n\sigma}(y)$ are the BdG wave functions found by solving the BdG equation corresponding to $\hat H_{\textrm{D}}$. 
To obtain the transmission amplitude, we apply the scattering matrix formalism \cite{Zocher13,Haim15b}. The S-matrix is $S(\epsilon)=\mathbbold{1}-2\pi i W^\dagger\left[\epsilon\mathbbold{1}-H_D+i\pi WW^\dagger\right]^{-1}W$ where $\mathbbold{1}$ denotes the unit matrix and $W$ the coupling matrix obtained from $\hat H_{\textrm{T}}$. If we assume that for each $N_0$, tunneling proceeds only via the Bogoliubov quasiparticle with the lowest energy, we find, at zero temperature and at the Fermi level, that the transmission is given by Eq.~(\ref{eq:breit-wigner}) with tunneling amplitudes $\lambda_{\sigma \alpha}(N_0) = \lambda_{n_{\textrm{min}} \sigma \alpha}/2$ for $\alpha=L,R$.

\emph{Results.} Before discussing the results in detail let us note that the presence of the superconductor may influence the visibility of the interference. Indeed, when the superconductor acts as a normal conductor, its level spacing is small, and we expect a rather large suppression of the interference and weak localization correction to the conductance~\cite{Huibers98}. Our main results for realistic parameter regimes \cite{Haim15a} are shown in Fig.~\ref{fig:results}. In the normal phase, the wire shows phase lapses when tunneling proceeds through the same level, Fig.~\ref{fig:results}(a). When tunneling occurs through consecutive levels, phases lapses are partially absent. In the trivial regime, superconductivity does not change the pattern. We now consider the topologically non-trivial regime. When electrons tunnel through the same level, we can distinguish the normal conducting and the superconducting regime by the presence or absence of phase lapses. For tunneling through consecutive levels phase lapses are always absent.

Interestingly, the transmission is sensitive to the spin polarizations of the MZMs at the ends of the wire. Defining $E_{\text{SO}} = m\alpha_R^2/(2\hbar^2)$, we find that, in the regime $B_z \gg E_{\text{SO}}$, both ends have the same spin polarization, in the $z$-direction. For $B_z \ll E_{\text{SO}}$, the ends have orthogonal polarizations along the $y$-axis, so electrons in the two arms of the interferometer will have orthogonal polarizations, leading to a suppression of the interference signal.

Note that the effective tunneling matrix elements alternate between $\lambda_{n\alpha}^u$ and $\lambda_{n\alpha}^v$ when summing over $N_0$. Furthermore the BdG wave functions, of a Hamiltonian invariant to the spatial symmetry operation $(x\rightarrow -x)$ with respect to the wire's mid-point, are either symmetric or anti-symmetric under this operation, such that only either the even or odd elements in the sums (\ref{eq:overlapu}) and (\ref{eq:overlapv}) are non-zero. We therefore expect the existence of phase lapses to be connected with the inversion (anti-)symmetry of BdG wave functions. Moreover the Hamiltonian (\ref{eq:wireBdG}) is invariant under the inversion symmetry $\Pi=\tilde{\Pi}\tilde{T}$. This implies that $u_{n\uparrow}(y)$ and $v_{n\downarrow}(y)$ behave in the same way under inversion and likewise for $u_{n\downarrow}(y)$ and $v_{n\uparrow}(y)$. On the other hand $u_{n\uparrow}(y)$ and $u_{n\downarrow}(y)$, and $v_{n\uparrow}(y)$ and $v_{n\downarrow}(y)$ behave in an opposite way under inversion. Naively we would then expect no phase lapses, regardless of whether the wire is in the topological or trivial regime. In the latter case however, the dominant particle and
hole like processes have opposite spin, such that products like $\lambda_{n\uparrow L}^u\lambda^{u \ast}_{n\uparrow R}$ and $\lambda_{n\downarrow L}^v\lambda^{v \ast}_{n\downarrow R}$ with the same sign dominate the transmission, and phase lapses occur. This is unlike the topological regime where the spins are mostly polarized.

We now study the transmission phase in the topological regime for the case where the BDI time-reversal symmetry is broken by a magnetic field, $B_x\neq 0$\cite{Osca14,Rex14,Nijholt16}. We consider a homogeneous wire longer than the localization length of the Majorana wave functions, in the regime $B_x \ll E_{\textrm{SO}} \ll B_z$. Here we can linearize the Hamiltonian (\ref{eq:wireBdG}) and find \cite{supplement}
\begin{align}
    \mathcal{\tilde{H}} =
    -i\hbar v_F \partial_y s_z\tau_z + \eps s_z + \tilde\Delta \tau_x s_z,
\label{eq:linearizedBdG}
\end{align}
acting on BdG wave functions $\Phi=[u_1,u_2,v_2,v_1]^T$ where $u_1,v_1$ and $u_2,v_2$ correspond to right- and left-movers. The Pauli matrices $s_\mu$ operate on these, while $\tau_\mu$ act on particles and holes.
They have opposite velocities $\pm v_F$, but the magnetic field along the wire axis causes an energy shift $\eps$. The effective pairing $\tilde\Delta$ is obtained from the relation $\eps / \tilde\Delta = B_x / \Delta$. Since a hard-wall boundary reflects left- into right movers, we obtain  $u_1(0)=-u_2(0)$ and $v_1(0)=-v_2(0)$. Then, the Majorana solution localized at the left end of the wire is given by $\Phi_L(y) \propto (1,-1,-ie^{-i\delta},ie^{-i\delta})^T e^{-(y/\hbar v_F)\sqrt{\tilde\Delta^2 - \eps^2}}$ with phase $\delta = \arcsin(\eps/\tilde \Delta)$. Identifying the tunneling amplitude as the component $u_1(0)$, and fixing its phase by requiring particle-hole symmetry, we thus find, at resonance,
\begin{align}
    \arg(T_{\up\up}) = \arcsin(B_x/\Delta) + \pi\cdot\text{integer},
    \label{eq:phaseEx}
\end{align}
which is independent of the total wire length.
This equation implies that upon breaking BDI symmetry, the interference pattern will not be at an extremum at zero flux. The phase shift will however will be identical for consecutive peaks.
To understand the phase shift $\arcsin(B_x/\Delta)$, we use an anti-unitary reflection symmetry of the Hamiltonian (\ref{eq:wireBdG}) and denote the BdG wave functions by $\Phi(y)=[u_\up,u_\down,v_\down,-v_\up]^T$. The left Majorana solution satisfies $\mathcal{H}\Phi_L=0$ and $P \Phi_L=\Phi_L$, such that $\Phi_R(y) = \Phi_L(L-y)^*$ is the Majorana solution at the right end. For a finite but sufficiently long wire, these will hybridize slightly to give the lowest energy Bogoliubov quasiparticle $\Phi \propto (\Phi_L \pm i\Phi_R)$, where the relative prefactor is fixed to $\pm i$ because the coupling Hamiltonian has particle-hole symmetry. Thus, the transmission amplitude is fully determined by the left Majorana solution, $T_{\up\up} \propto u_\up(y_L)u_\up^*(y_R) = \mp i u_{\up,L}(y_L)^2$ and does not depend on the wire length if the BdG equation does not explicitly depend on it.

\emph{Conclusion.} We have studied the transmission of electrons through a hybrid system consisting of a quantum wire and an s-wave superconductor in the Coulomb blockade regime in a magnetic field and with strong spin-orbit coupling. In this regime the system supports zero-energy MZMs. We consider only tunneling through a single Bogoliubov quasiparticle. In this case we found that the existence of phase lapses in the transmission phase depends on the superconducting gap and on whether the wire is in the topological or trivial phase. We expect that it should be possible to measure this effect in interference experiments.

\emph{Acknowledgments.} B.R. acknowledges financial support from grant
SFB 762 and DFG RO 2247/8-1.  A.S and Y.O. acknowledges financial support by the Israeli Science Foundation (ISF), the ERC Grant agreement No. 340210, DFG CRC TR 183, a BSF, and Microsoft station Q grant. We thank R. Lutchyn and  C. Marcus for fruitful discussions.

\bibliographystyle{prl}
%\newpage

%%%%%% Supplementary Material %%%%%%%%

\clearpage

\widetext

\begin{center}
\textbf{\large Supplemental Material}
\end{center}

\setcounter{equation}{0}
\setcounter{figure}{0}
\setcounter{page}{1}
\makeatletter
\renewcommand{\theequation}{S\arabic{equation}}
\renewcommand{\thefigure}{S\arabic{figure}}
\renewcommand{\bibnumfmt}[1]{[S#1]}
\renewcommand{\citenumfont}[1]{S#1}

\section{Derivation of the transmission from the S-matrix}

For completeness we here show how to derive Eq.~(\ref{eq:breit-wigner}) of the main text. We thus assume a two-level system with Hamiltonian

\begin{align}
H=\left(\begin{array}{cc} \epsilon_1 & 0 \\ 0 & \epsilon_2 \end{array}\right),
\end{align}
with eigenenergies $\epsilon_1$ and $\epsilon_2$. The system described by this Hamiltonian is taken to be coupled to two leads. The coupling is described by the coupling matrix

\begin{align}
W=\left(\begin{array}{cc} \lambda_{1L} & \lambda_{1R} \\ \lambda_{2L} & \lambda_{2R} \end{array} \right),
\end{align}
where we use the same notation as in the main text. From this we can form the S-matrix using

\begin{align}
\label{eq:weid}
S(\epsilon)=\mathbbold{1}-2\pi i W^\dagger\left[\epsilon\mathbbold{1}-H+i\pi WW^\dagger\right]^{-1}W.
\end{align}
The S-matrix describes the relationship between incoming and outgoing states scattering of a system, here described by $H$. From it, we can now read of the transmission as the off-diagonal element. In order to arrive at our result in Eq.~(\ref{eq:breit-wigner}), we replace the eigenenergies with the single-particle energies discussed in the main text. To gain some intuition into the appearance of phase lapses we then calculate the transmission in the limit of weak couplings. In this case, we can neglect off-diagonal terms in the matrix in the square brackets of (\ref{eq:weid}), making the matrix inversion straight-forward. In order to estimate the magnitude of corrections, we finally Taylor expand in the off-diagonal terms and replace the denominator by its value $eV_G/2$ in between resonances. We hence arrive at Eq.~(\ref{eq:breit-wigner}).

This approximation is accurate at the vicinity of the resonances \cite{supHackenbroich00}. In between resonances the off-diagonal terms may determine the width of the phase lapse. Taking into account only two resonances, and focusing on the large $B$ limit where the spins are polarized,  we can extract the transmission matrix element of the scattering matrix to be,
\begin{equation}
\frac{1}{{\rm Det}} \left(\lambda_{1L}^*, \lambda_{2 L}^* \right) \left( \begin{array}{cc} \epsilon-\epsilon_2 +i \left( |\lambda_{2L}|^2+ |\lambda_{2R}|^2 \right) & -i
\left( \lambda_{1L} \lambda_{2L}^*+ \lambda_{1R} \lambda_{2R}^* \right) \\
-i \left( \lambda_{1L}^* \lambda_{2L}+ \lambda_{1R}^* \lambda_{2R} \right) & \epsilon-\epsilon_1 +i \left( |\lambda_{1L}|^2+ |\lambda_{1R}|^2 \right)
\end{array} \right) \left( \begin{array}{c} \lambda_{1R} \\ \lambda_{2R}\end{array} \right).
\label{transmission_app}
\end{equation}
Here $\rm Det$ is the determinant of the $2 \times 2$ matrix in (\ref{transmission_app}). Away from a resonance this determinant is not singular, and a sharp phase lapse may occur only as a consequence of the variation of the numerator with energy. More specifically, a sharp phase lapse occurs if the numerator (and hence the transmission) passes through zero. The expression (\ref{transmission_app}) may be simplified to be
\begin{equation}
\frac{1}{{\rm Det}}\left(\lambda_{2L}^*(\epsilon-\epsilon_1)\lambda_{2R}+\lambda_{1L}^*(\epsilon-\epsilon_2)\lambda_{1R}\right)
\label{transsimple}
\end{equation}
As is evident from Eq. (\ref{transsimple}), a particular phase relation between the different tunneling matrix elements is needed for the transmission to vanish and go through a sharp $\pi$-lapse of its phase. In particular, a sharp phase lapse occurs between the two resonances when $\arg{\lambda_{1L}^*\lambda_{1R}}=\arg{\lambda_{2L}^*\lambda_{2R}}$.

\section{Absence of phase lapses in the presence Majorana end states}

We here show that phase lapses do not occur if the wire supports Majorana zero modes. To do this we assume that tunneling proceeds through the (almost) zero energy Bogoliubov quasiparticle state $\beta_0$, whose electron operator is $c_0(y) = u(y)\beta_0 + v^*(y)\beta_0^\dagger=\xi_L(y)\gamma_1 + \xi_R(y)\gamma_2$,  with associated Majorana operators $\gamma_1,\gamma_2$ and  Majorana wave functions $\xi_L(y),\xi_R(y)$ localized near the left lead $y_L$ and right lead $y_R$.
Although in principle the superconductor accommodates only pairs of electrons, $N=2M$, the zero energy quasiparticle state can accommodate one extra electron, thus allowing odd $N=2M+1$ as well. When $N=2M$, the quasiparticle state is unoccupied, and it is plausible that an additional electron tunnels via the particle-like part of the Bogoliubov-de Gennes wave function, $\lambda_{\sigma L,R}(2M) \propto u_\sigma(y_{L,R};2M)$. On the other hand, when $N=2M+1$, the quasiparticle is occupied and needs to be emptied, so that the hole-like part is relevant for tunneling $\lambda_{\sigma L,R}(2M+1) \propto v^*_\sigma(y_{L,R};2M+1)$. If we further assume that the quasiparticle wave function is the same for different particle numbers, $u_\sigma(y;2M)\approx u_\sigma(y,2M+1)$ and similarly for $v$, then the phases of consecutive resonances satisfy
\begin{align}
\label{eq:phaselapse}
    \frac{\lambda_{\sigma L}(2M+1)\lambda^*_{\sigma R}(2M+1)}{\lambda_{\sigma L}(2M)\lambda^*_{\sigma R}(2M)}
    \approx \frac{u_{\sigma}(y_L)u_{\sigma}(y_R)^*}{v^*_{\sigma}(y_L)v_{\sigma}(y_R)}
    = -1.
\end{align}
The last equality follows from the localization property of the Majorana wave functions $\xi_L,\xi_R$, as well as from the requirement that $\gamma_1,\gamma_2$ be Hermitian, and that $\beta_0,\beta_0^\dagger$ satisfy the canonical anticommutation relations. This establishes the absence of phase lapses when the wire supports Majorana zero modes.

\section{Effect of spin}

To estimate the effect of spin we express the zero energy solution as a superposition of three decaying waves ($e^{-\kappa y}$ with $\Re \kappa > 0$): one evanescent wave ($\Im\kappa=0$), and two oscillating waves $(\Im\kappa = \pm k_F)$. The decay of the evanescent wave is mostly determined by the Zeeman gap $B_z/E_{\text{SO}}$, whereas the decay of the oscillating waves is determined by the superconducting gap $\Delta/E_{\text{SO}}$. In the regime $B_z \gg E_{\text{SO}}$, the oscillating waves dominate, in contrast to the case $B_z \ll E_{\text{SO}}$. This ultimately leads to the net spin polarizations discussed in the main text.

\section{Linearization of the Hamiltonian with broken time-reversal symmetry}

We consider the regime $B_x \ll E_{\textrm{SO}} \ll B_z$ and linearize the Hamiltonian given in Eq.~(\ref{eq:wireBdG}) of the main text, reproduced here for convenience
\begin{align}
\mathcal{H} &=
\left[-\frac{\hbar^2}{2m}\partial_y^2-\mu-i\alpha_R\sigma_x\partial_y\right]\tau_z
- B_z\sigma_z - B_x\sigma_x + \Delta\tau_x
.\end{align}

In the absence of superconductivity, this Hamiltonian has two bands.
We first project to the lower band with dispersion relation
$E_-(k) = \hbar^2k^2/2m-\mu - \sqrt{B_z^2 + (\alpha_R k - B_x)^2}$.
If we denote the corresponding eigenspinor by $\phi_-(k)$, we can implement the projection by introducing a fermionic field $\psi_0(k)$ and making the ansatz $(\psi_\up(k),\psi_\down(k)) = \phi_-(k)\psi_0(k)$.
Reintroducing superconductivity, the projected second-quantized Hamiltonian is
$\hat H_{\text{proj.}} = \sum_k [E_-(k)\psi_0^\dagger(k) \psi_0(k) + \Delta_-(k)\psi_0(-k)\psi_0(k) + \text{h.c.}]$ where $\Delta_-(k)$ is now an effective $p$-wave pairing.
At $B_x=0$, it can be expressed as $\Delta_-(k) = (\Delta/2)\partial E_-(k)/\partial B_x$
thanks to the identity $\partial E_-(k)/\partial B_x=-\langle\phi_-(k)|\sigma_x|\phi_-(k)\rangle$.
We now linearize about two fixed momenta $\pm k_0$ by expanding the field $\psi_0$ into right- and left-moving fields $\psi_1,\psi_2$ with respect to these momenta. To first order, the Zeeman energy $B_x$ will lead to a shift of the energy of the left-movers relative to the right-movers.
We keep only this contribution and discard any other contribution beyond zeroth order. We obtain a linearized BdG Hamiltonian
\begin{align}
    \mathcal{\tilde{H}} =
    -i\hbar v_F \partial_y s_z\tau_z + \eps s_z + \tilde\Delta \tau_x s_z
,\end{align}
with second quantized form $\tilde H = \frac12 \int dy\, \Psi^† \widetilde{\mathcal{H}} \Psi$ with $\Psi=(\psi_1, \psi_2, \psi_2^\dagger, \psi_1^\dagger)^T$. Here, the Pauli matrices $s_\mu$, act on right- and left-movers, while $\tau_\mu$ act on particles and holes.
The Fermi velocity is $v_F=(1/\hbar)(\partial E_-/\partial k)(k_0)$, the relative energy shift is $\eps = B_x(\partial E_-(k_F)/\partial B_x)$, while the new effective order parameter has the form $\tilde\Delta = \Delta(\partial E_-(k_F)/\partial B_x)$.

\section{Details of numerics}

\begin{figure}[h]
\includegraphics[width=0.65\linewidth]{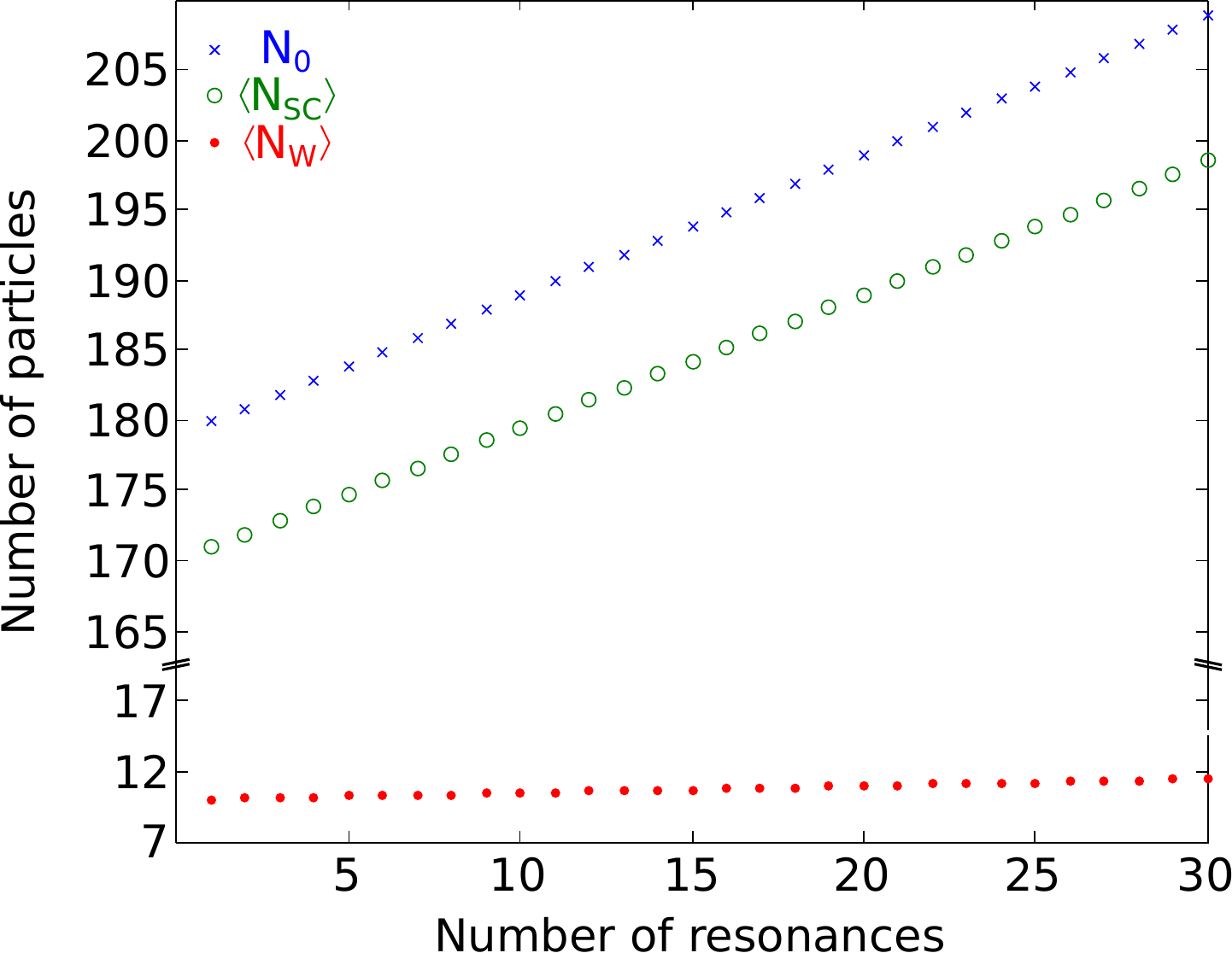}
\caption{Number of particles in the system, calculated such that the wire is in the topological regime. The number of particles in the wire and superconductor respectively are found by minimizing Eq.~(\ref{eq:chargingham}) of the main text with respect to $N_{\textrm{w}}$. The crosses show the total number of particles in the system, the circles the number of particles in the superconductor, whilst the dots show the number of particles in the wire. The calculations for subsequent resonances have been done assuming $E_{\textrm{M}}/E_{\textrm{w}}=0.05$ such that every 20th electron added to the system is added to the wire.}
\label{fig:partnumbers}
\end{figure}

In order to solve the BdG equations we discretize the BdG equations on a one-dimensional lattice with $500$ sites. To solve these discretized equations it is necessary to determine the chemical potential corresponding to a given particle number in the wire. As discussed in the main text, the latter is obtained by minimizing the charging energy and is given by $N_{\textrm{w}}=(E_{\textrm{M}}/E_{\textrm{w}})N_0$, where $E_{\textrm{w}}$ is the charging energy in the wire, and $E_{\textrm{M}}$ the coupling. $N_0$ denotes the total number of particles in the system. We now write the wire number operator in terms of BdG operators and take the expectation value with respect to the ground state. This yields the number of particles in the wire, in the ground state

\begin{align}
\label{eq:groundstatepart}
N_{\textrm{w}}=\sum_{n,\sigma}\int dy\,|v_{n\sigma}(y)|^2,
\end{align}

where the sum is taken over states with positive energy. If now a quasiparticle is added through the Bogoliubon with lowest energy, the total particle number is changed by

\begin{align}
\label{eq:partdiff}
\delta N=\sum_\sigma \int dy\,\left[|u_{n_{\textrm{min}}\sigma}(y)|^2-|v_{n_{\textrm{min}}\sigma}(y)|^2\right],
\end{align}

where we recall that $n_{\textrm{min}}$ labels the BdG quasiparticle with the lowest energy. In Eqs.~(\ref{eq:groundstatepart}) and (\ref{eq:partdiff}), $u_{n\sigma}(y)$ and $v_{n\sigma}(y)$ are BdG wave functions, which depend on the chemical potential through the BdG Hamiltonian. By fixing the number of particles in the wire, such that the wire is either in the topological or trivial regime, we can thus obtain the chemical potential self-consistently. The number of particles used for the calculations in the topological regime are shown in Fig.~\ref{fig:partnumbers}.

\section{Derivation of the effective tunneling Hamiltonian}

We here show how to derive the tunneling Hamiltonian given in Eq.~(\ref{eq:efftun}) of the main text. The starting point is the tunneling Hamiltonian (\ref{eq:tunhamorig}) reproduced here for completeness
\begin{align}
\hat H_{\textrm{T}} &= \sum_{\substack{m\sigma \\ \alpha=L,R}}t_{\alpha m\sigma}c_\sigma^\dagger(y_\alpha)d_{m}+{\textrm{h.c.}}.
\end{align}
Recall that, $\alpha$ labels the left and right leads respectively, $\sigma$ denotes spin, $c_\sigma(y_\alpha)$, $c_\sigma^\dagger(y_\alpha)$ are lead operators evaluated in the vicinity of lead $\alpha$, $d_{m}$ ($d_{m}^\dagger$) annihilates (creates) an electron in the $m$'th wire state with energy $\epsilon_m$, and $t_{\alpha m\sigma}$ is the tunneling matrix element between lead $\alpha$ and the $m$'th wire state. The wire operators $d_m$ can now be expressed in terms of the local fermionic operator $d_\sigma(y)$ and the normal-state BdG eigenfunctions $\varphi_{m\sigma}(y)$ as
\begin{align}
d_{m}=\sum_\sigma \int dy\,\varphi_{m\sigma}^\ast(y)d_\sigma(y),
\end{align}
In the presence of the pairing term, we can expand the fermionic operator $d_\sigma(y)$ in terms of BdG quasiparticles such that

\begin{align}
d_\sigma(y)=\sum_ne^{-i\phi/2}\left[u_{n\sigma}(y)\beta_n+v_{n\sigma}^\ast(y)\beta_n^\dagger\right].
\end{align}
Here, $e^{-i\phi/2}$ lowers the total charge $\hat N_0$ by one, and $u_{n\sigma}(y)$ and $v_{n\sigma}(y)$ are the BdG wave functions found by solving the BdG equation corresponding to $\hat H_{\textrm{D}}$.
With this, the tunneling part of the Hamiltonian becomes
\begin{align}
\hat H_{\textrm{T}} = \sum_{\substack{\alpha=L,R \\ n\sigma}}
    c_\sigma^\dagger(y_\alpha) e^{-i\phi/2}
    [\lambda^u_{n\sigma\alpha} \beta_n + \lambda^v_{n\sigma\alpha} \beta_n^\dagger]
    +\textrm{h.c.},
\end{align}
where the matrix elements
\begin{align}
\lambda_{n\sigma\alpha}^u &= \sum_m t_{\alpha m\sigma}\int dy\, \varphi_{m\sigma}^\ast(y)u_{n\sigma}(y), \\
\lambda_{n\sigma\alpha}^v &= \sum_m t_{\alpha m\sigma}\int dy\, \varphi_{m\sigma}^\ast(y)v_{n\sigma}^\ast(y).
\end{align}
describe the coupling of the particle- and hole-like parts to the leads.

As discussed in the main text, we now assume that the gap in the superconductor is greater than in the wire. Thus for $N_0$ odd, the unpaired electron goes into the wire and occupies the lowest lying BdG quasiparticle state. We hence choose the creation and annihilation operators such that $2\beta_{n_{\textrm{min}}}^\dagger\beta_{n_{\textrm{min}}}-1=(-1)^{N_0}$ in the ground state. The other quasiparticle states are not occupied, so $\beta_n^\dagger\beta_n=0$ is zero in the ground state for $n\neq n_{\textrm{min}}$. Following Ref.~\cite{supFu10}, we can project the Hamiltonian to a tunneling problem through a single resonant level. For that, we project the Hamiltonian to the states with $N_0$ and $N_0+1$ particles, and set $s_{\pm}=e^{\pm i\phi/2}$. We then introduce auxiliary Majorana operators such that
\begin{align}
\beta_n=\frac{1}{2}\left(\gamma_{b1}^{(n)}-i\gamma_{b2}^{(n)}\right),
\end{align}
and map
\begin{align}
\gamma_{b1}^{(n_{\textrm{min}})}s_{-}&\rightarrow f_{n_{\textrm{min}}}, \\
\gamma_{b1}^{(n_{\textrm{min}})}s_{+}&\rightarrow f_{n_{\textrm{min}}}^\dagger, \\
\gamma_{b2}^{(n_{\textrm{min}})}s_{-}&\rightarrow -i(-1)^{N_0}f_{n_{\textrm{min}}}, \\
\gamma_{b2}^{(n_{\textrm{min}})}s_{+}&\rightarrow i(-1)^{N_0}f_{n_{\textrm{min}}}^\dagger.
\end{align}
The new operators satisfy fermionic anticommutation relations, and the dependence on $N_0$ ensures that $\langle f_{n_{\textrm{min}}}^\dagger f_{n_{\textrm{min}}}\rangle = 0$ in the ground state. A similar mapping can be performed for $n\neq n_{\textrm{min}}$, with the caveat that the occupation of quasiparticles $\beta_n^\dagger\beta_n$ is always zero in this case, such that the dependence on $N_0$ vanishes. The tunneling part thus becomes
\begin{align}
H_{\textrm{T}}=\frac{1}{2}\sum_{\substack{\alpha=L,R \\ n\sigma}}\lambda_{n\sigma\alpha}c_\sigma^\dagger(y_\alpha)f_n+\textrm{h.c.},
\end{align}
with effective tunneling matrix elements
\begin{align}
\lambda_{n_{\textrm{min}}\sigma\alpha}=\lambda_{n_{\textrm{min}}\sigma\alpha}^u+\lambda_{n_{\textrm{min}}\sigma\alpha}^v - (-1)^{N_0}\left[\lambda_{n_{\textrm{min}}\sigma\alpha}^u-\lambda_{n_{\textrm{min}}\sigma\alpha}^v\right]
,\end{align}
and $\lambda_{n\sigma\alpha}=2\lambda_{n\sigma\alpha}^u$ for $n\neq n_{\textrm{min}}$. This is the effective tunneling Hamiltonian quoted in the main text.

\bibliographystyle{prl}

\begin{thebibliography}{99}
\bibitem{lapse_detection} Notice that the presence of the phase lapse can be detected, even without measuring exactly at what value of $V_G$ it occurs. This can be done by comparing the phase shift in the Aharonov-Bohm oscillations at the descending part of the N'th peak and the ascending part of the $(N+1)$'th peak. If they are shifted by $\pi$, an odd number of phase lapses occurs between them.
\end{thebibliography}
%\newpage

\end{document}